\documentclass[preprints,article,accept,moreauthors,pdftex,10pt,a4paper]{mdpi}

\firstpage{1} 
\pubvolume{xx}
\issuenum{1}
\history{Received: date; Accepted: date; Published: date}

\usepackage{undertilde}

\DeclareMathOperator{\Tr}{Tr}

\newcommand{\Mv}[1]{\langle #1 \rangle}

\newcommand{\braket}[2]{\langle #1 | #2 \rangle}

\newcommand{\ket}[1]{| #1 \rangle}
\newcommand{\bra}[1]{\langle #1 |}

\newcommand{\Tlambda}{\utilde{\lambda}}
\newcommand{\me}{\mu^E}

\Title{New equilibrium ensembles for isolated quantum systems}

\Author{Fabio Anza $^{1}$\orcidA{}}

\AuthorNames{Fabio Anza}

\address{%
$^{1}$ \quad Clarendon  Laboratory,  University  of  Oxford,  Parks  Road,  Oxford  OX1  3PU,  United  Kingdom 1; fabio.anza@physics.ox.ac.uk}

\corres{Correspondence: fabio.anza@physics.ox.ac.uk;}

\abstract{The unitary dynamics of isolated quantum systems does not allow a pure state to thermalize. Because of that, if an isolated quantum system equilibrates, it will do so to the predictions of the so-called ``diagonal ensemble'' $\rho_{DE}$. Building on the intuition provided by Jaynes' maximum entropy principle, in this paper we present a novel technique to generate progressively better approximations to $\rho_{DE}$. As an example, we write down a hierarchical set of ensembles which can be used to describe the equilibrium physics of small isolated quantum systems, going beyond the ``thermal ansatz'' of Gibbs ensembles.}


\begin{document}



\section{Introduction}

The theory of Statistical mechanics is meant to address the equilibrium physics of macroscopic systems. Both at the classical and at the quantum level, the whole theory is based on the \emph{thermal equilibrium} assumption. Such physical condition is expressed, mathematically, by saying  that the system is described by one of the so-called Gibbs ensembles\cite{Greiner1995,Schrodinger1989}. The theory has enjoyed a marvelous success in explaining and predicting the phenomenology of large quantum systems and, nowadays, we use its results and tools well beyond the domain of physics. However, strictly speaking, this course of action is justified only in the thermodynamic limit. More realistically, the assumptions of Statistical Mechanics are believed to be justified on a scale of, say, an Avogadro number of particles $N \sim 10^{23}$. Despite that, both theoretical analysis\cite{Srednicki,Srednicki1999,Srednicki1996,Deutsch1991a,Rigol2008} and experimental investigations \cite{Gring2012,Trotzky2012,Pertot2014} suggest that Statistical Mechanics is able to describe equilibrium phenomena even in small isolated quantum systems\cite{Polkovnikov2016}. In turn, this points towards and ``early emergence'' of the thermal equilibrium assumption, already for systems of modest sizes.

This picture results from a large-scale effort of the scientific community to understand the thermalization mechanisms and provide solid foundations to the emergence of Statistical Mechanics\cite{Reimann2010,Polkovnikov2011b,DAlessio2016,Eisert2015a,Reimann,Reimann2015,Reimann2016,Reimann2012,Balz2017,Torres-Herrera2016,Torres-Herrera2015,Borgonovi2016,Santosa,Santos,Borgonovi2017,Santos2012,Santosb,Borgonovi2017a,Gogolin2016}. Borrowing the terminology from Seth Lloyd's PhD thesis \cite{Lloyd1988}, we put all these works under the name of ``Pure States Quantum Statistical Mechanics''. The theory is not yet a coherent and well understood set of statements, but it is founded on four main approaches: the Quantum Chaos approach\cite{Borgonovi2016,Torres-Herrera2016,Torres-Herrera2015,Santosa,Santos}, the Eigenstate Thermalisation Hypothesis (ETH) \cite{Mori2018,Reimann2015a,DAlessio2016,Anza2018,Anza2017,Anza2018a}, the so-called Typicality Arguments \cite{Reimannb,Gemmer2010,Goldstein2010,Goldstein2010a,Goldstein2006}  and the Dynamical Equilibration Approach\cite{Gogolin2016}. All these approaches have a highly non-trivial overlap and their interplay is not yet fully understood.

Moreover, in the last ten years, the will to provide solid foundations to statistical mechanics met the necessity to understand how thermodynamics is modified at the nanoscale, where size-dependent fluctuations and quantum effects are not negligible \cite{Goold2016,Klages2013,Jarzynski2011,Seifert2012,Kosloff2013}. In this work, we use this mindset and apply it to the equilibrium physics of (small) isolated quantum systems. Building on Jaynes' Maximum Entropy Principle\cite{Jaynes1957,Jaynes1957a}, we develop a novel technique to generate progressively better approximations to the equilibrium state of an isolated quantum system. Using this perspective, the Canonical Gibbs Ensemble is understood as the first-level of a hierarchical set of ensembles which can describe, in a progressively more precise manner, the equilibrium behaviour of isolated quantum systems. A similar point of view was previously offered, in connection with the dynamics of classically integrable quantum systems, in Ref.\cite{Sels2015,Kim2017}.


The relevance of our work stems from two main points of view. On the one hand, we are interested in a better understanding of the equilibrium physics of mesoscopic and microscopic isolated quantum systems. In particular, we are interested in developing a theory for isolated quantum systems at equilibrium which goes beyond the thermal assumption. In turn, this will allow us to improve our understanding of the conditions which lead to the emergence of thermal equilibrium. On the other hand, because of the large domain of use of Thermodynamics and Statistical Mechanics, the relevance of this issue goes beyond the realm of physics. For example, Thermodynamical statements are ordinarily made in Biology and Biochemistry. Relying on such statements implicitly assumes the validity of some ``thermodynamic limit'' which might not be justified at the small scale. For this reason, we ask: How does the picture change when both quantum effects and system-size-dependent fluctuations are not negligible? Our framework can be used to answer this and similar questions. \\

\section{Isolated quantum systems}

Throughout the paper, we always deal with quantum systems which can be described with an Hilbert space $\mathcal{H}$ of finite dimensions $D$. Moreover, we assume to deal with a modular system made by $N$ identical units, each described by an Hilbert space of dimension $d$. Hence, $D$ grows exponentially with the size $N$ of the system: $D \sim d^N$. Now we take a step back and look at the unitary dynamics of isolated quantum systems, assuming it is generated by a non-degenerate and time-independent Hamiltonian $H$, with the following spectral decomposition: $H=\sum_{n=1}^D \epsilon_n \Pi_n$ and $\Pi_n := \ket{\epsilon_n} \bra{\epsilon_n}$. Given an initial state $\ket{\psi_0} \in \mathcal{H}$, the solutions of the dynamical problem are given by a one-parameter (time) family of states $\ket{\psi_t} = U(t)\ket{\psi_0}$, where $U(t)$ is the unitary propagator $U(t):=e^{-\frac{i}{\hbar}Ht}$. Thanks to the non-degeneracy assumption on $H$, its eigenbasis $\left\{ \ket{\epsilon_n}\right\}$ is unique and it provides a basis for the Hilbert space $\mathcal{H}$. Thus, given $\ket{\psi_0} = \sum_n c_n \ket{\epsilon_n}$, if we expand the time-dependent density matrix $\rho(t) := \ket{\psi_t}\bra{\psi_t}$ in the energy basis $\ket{\epsilon_n}$ 
\begin{equation}
\rho(t) = \underbrace{\sum_{n=1}^D  |c_n|^2 \, \ket{\epsilon_n} \bra{\epsilon_n}}_{\rho_{\mathrm{DE}}} + \underbrace{\sum_{n \neq m } c_n c^{*}_m e^{- \frac{i}{\hbar}(\epsilon_n - \epsilon_m)t} \ket{\epsilon_n} \bra{\epsilon_m} }_{\delta(t)}
\end{equation}
we can see that there are two distinct contributions. The first one ($\rho_{\mathrm{DE}}$) is called the \emph{diagonal ensemble} and it does not depend on time. This is also the state that we obtain after performing an infinite-time average
\begin{equation}
\lim_{T \to \infty} \frac{1}{T} \int_0^T \rho(s) \, ds = \rho_{\mathrm{DE}} + \sum_{n \neq m} c_n c_m^{*}\lim_{T \to \infty} \frac{1}{T} \int_0^T e^{-\frac{i}{\hbar}(\epsilon_n - \epsilon_m)s}ds = \rho_{\mathrm{DE}} + \sum_{n \neq m} c_n c_m^{*} \delta_{n,m} = \rho_{\mathrm{DE}}  \, \,\, .
\end{equation}
The second one ($\delta(t)$) accounts for the time-dependent fluctuations of $\rho(t)$ around $\rho_{\mathrm{DE}}$.

In the energy eigenbasis, the dynamics affects only the phases of the coefficients: $\braket{\epsilon_n}{\psi_t}= c_n e^{- \frac{i}{\hbar}\epsilon_n t}$, leaving their modulus unchanged. Thus, given any initial state $\ket{\psi_0}$, specified by its decomposition in the energy basis $\left\{ c_n=\braket{E_n}{\psi_0}\right\}$, the state will never forget completely about its initial conditions. There is always a $D-1$ number of (linearly independent) conserved quantities $|c_n|^2 := |\braket{\psi_0}{\epsilon_n}|^2$. They are the probabilities of finding the system in the eigenstate $\ket{\epsilon_n}$ after we measure the energy $H$ and $P_E := \left\{ p(\epsilon_n) := \Tr \rho(0) \Pi_n = |c_n|^2\right\}_{n=1}^D$ is the probability distribution of $\rho(0)$ over the eigenvalues of $H$. It can be easily seen that the whole probability distribution never changes in time. Because of that, as shown in Ref.\cite{Linden,Gogolin}, if an observable $A$ equilibrates, it will do so to the predictions of the diagonal ensemble $\rho_{\mathrm{DE}}$:
\begin{equation}
\Mv{A}_{DE} = \Tr A\rho_{DE} = \sum_{n=1}^D |c_n|^2 \bra{E_n} A \ket{E_n} \,\,\, .
\end{equation} 

\section{Results}\label{sec:Results}
There is a simple connection between $\rho_{DE}$ and Jaynes' Maximum Entropy Principle. The mindset is as follows. Suppose we are forced to make predictions about the state of a quantum system, given that we have some knowledge about its state, like the average value of the energy. Statistical Inference can be used to tackle this issue. The Maximum Entropy principle states that our best guess is the state that maximises the von Neumann entropy $S_{\mathrm{vN}}(\rho) := - \Tr \rho \log \rho$, compatibly with the presence of some constraints $\left\{\mathcal{C}_n = 0\right\}$ which represent our state of knowledge about the system. Since we are interested in addressing the equilibrium properties, it is natural to choose the full set $P_E$ of conserved quantities as constraints. While this is highly impractical, as it requires the knowledge of all energy eigenstates, it is the correct thing to do as $P_E$ provides a complete set of linearly independent conserved quantities. Therefore, given an initial state $\ket{\psi_0} = \sum_n c_n \ket{\epsilon_n}$,  our constraints will be $\mathcal{C}_n := \Tr \rho \Pi_n - |c_n|^2$. For each constraint $\mathcal{C}_n$ we introduce a Lagrange multiplier $\lambda_n$ and then define the auxiliary function $\Lambda (\rho,\left\{\lambda_n\right\}) = S_{\mathrm{vN}}(\rho) + \sum_{n=1}^D \lambda_n \mathcal{C}_n$, which can be freely optimized. The result yields the state $\rho_{DE}$ and the associated $D$ Lagrange multipliers $\left\{\lambda_n\right\}_{n=1}^D$ acquire a simple form $\lambda_n = 1 + \log |c_n|^2$. Here we are also assuming that all the $c_n$ considered are non-vanishing. This means that we are working with an effective subspace of the whole Hilbert space where we got rid of all the symmetries in the Hamiltonian. \\

We will now explore the consequences of the fact that the constraints $\left\{ \mathcal{C}_n\right\}$ are linear functionals of the density matrix. Because of that, using linear combinations of the $\mathcal{C}_n$ does not change the result of the optimization procedure. In other words, using the constraints $\left\{\mathcal{C}_n\right\}$ or linear combinations of them $\left\{\utilde{\mathcal{C}}_n\right\}$ does not change the solution. In formulas, for any non-singular matrix $M$, with inverse $M^{-1}$, the auxiliary function $\Lambda(\rho,\left\{ \lambda_n\right\})$ has the following symmetry
\begin{equation}
\Lambda(\rho,\left\{ \lambda_n\right\}) = S_{vN}(\rho) + \sum_{k} \lambda_k \mathcal{C}_k = S_{vN}(\rho) + \sum_{n} \utilde{\lambda}_n \utilde{\mathcal{C}}_n = \utilde{\Lambda} (\rho,\{ \utilde{\lambda}_n\})\label{eq:rot}
\end{equation}
with $\utilde{\mathcal{C}}_n = \sum_{k} M_{nk} \mathcal{C}_k$ and  $\utilde{\lambda}_n = \sum_{k} \lambda_k (M^{-1})_{kn}$. As long as the transformation $M$ is non-singular, it can be absorbed in the value of the Lagrange multipliers, applying the inverse transformation $M^{-1}$. This invariance under linear transformation can, and should, inspire different approximation schemes. In principle, the optimization problem should include the set of all conserved quantities $\left\{ \Tr \rho \Pi_n \right\}$. However, this is highly impractical, as it requires the exact knowledge of all energy eigenstates and their overlaps with the initial state. Nevertheless, depending on the physical situation under scrutiny, one might be able to choose decent approximation schemes to obtain a solution which is sufficiently accurate for the required purposes. Thinking about entropy maximization in a geometric fashion, the entropy landscape is a manifold which can be parametrized with different coordinate systems. In other words, in the convex set of the density matrices, we are using different coordinates to describe the very same landscape. Thus, it is natural to expect that a ``suitable'' choice of coordinates, adapted to the landscape, can inspire useful approximation schemes. 

 In fact, this is what we have already been doing in using the thermal ansatz for isolated quantum systems at equilibrium. To see this, we choose a specific form for the matrix $M$: the Vandermonde matrix \cite{Horn1994,Macon1958} of all the Hamiltonian eigenvalues, $M_{hk}=\left(V_H\right)_{hk} := \left(E_k\right)^{h-1}$. This is non-singular as its determinant is $\det M = \prod_{1 \leq n \le k \leq D} (E_n - E_k)$ and we already assumed to deal with a non-degenerate Hamiltonian. With such choice for the matrix $M$ we have
\begin{equation}
\utilde{\mathcal{C}}_n = \sum_{k}(V_H)_{nk}\mathcal{C}_k = \Tr \rho H^{n} - \sum_{j}|c_j|^2 E_j^{n} = \Tr \rho H^{n} - \me_{n} \,\,\, . 
\end{equation}
From the equivalence in Eq.(\ref{eq:rot}) we can see that using the full set of conserved quantities $P_E$ is equivalent to using the set of the $D$ statistical moments $\mu_j^E:=\Tr \rho_{DE} H^j = \sum_{k=1}^D |c_k|^2 E_k^j$ of the energy probability distribution $P_E$. Hence, $\utilde{\mathcal{C}}_0$ is the normalization of the state $\Tr \rho -1=0$; $\utilde{\mathcal{C}}_1$ accounts for the fixed average value of the energy $\Tr \rho H - \me_1=0$; $\utilde{\mathcal{C}}_2$ for the fixed average value of the square Hamiltonian $\Tr \rho H^2 - \me_2=0$ and so on. Such way of writing the constraints inspires an approximation scheme that we already know: If we have no information whatsoever about the energy of the system, we use as constraint only the normalization of the state $\utilde{\mathcal{C}}_0=0$. The result of the optimization procedure gives the \emph{first-level ensemble}, which we call $\gamma_0$. From the experimental perspective, this corresponds to the situation in which the energy fluctuates so much that it is meaningless to trust its first moment.
\begin{equation}
\gamma_0 = e^{(\utilde{\lambda}_0 -1)\mathbb{I}} = \frac{\mathbb{I}}{\mathcal{Z}_0} \qquad \qquad \qquad  \utilde{\lambda_0} = 1 - \log \mathcal{Z}_0  \qquad \qquad \qquad  \mathcal{Z}_0 = D
\end{equation}
If the fluctuations are not too wild, we can trust the first moment of $P_E$ to give meaningful information about the system. Therefore, the constrained optimization of the von Neumann entropy is performed compatibly with the presence of the first two constraints $\utilde{\mathcal{C}}_0=\utilde{\mathcal{C}}_1=0$. This gives the \emph{first-level ensemble} $\gamma_1$: 
\begin{equation}
\gamma_1 = e^{(\utilde{\lambda}_0 - 1)\mathbb{I}} e^{(\utilde{\lambda}_1 -1) H} = \frac{e^{(\utilde{\lambda}_1 -1) H}}{\mathcal{Z}_1(\utilde{\lambda}_1)}   \qquad \qquad\qquad \utilde{\lambda}_0 = 1- \log \mathcal{Z}_1(\utilde{\lambda}_1) \qquad \qquad  \frac{\partial \log \mathcal{Z}(\utilde{\lambda}_1)}{\partial \utilde{\lambda}_1} = \me_1
\end{equation}
where $\mathcal{Z}_1(\utilde{\lambda}_1):=\Tr e^{(\utilde{\lambda}_1 -1) H}$. Here we recognize the Canonical Gibbs' Ensemble:
\begin{equation}
\gamma_1 = \rho_G(\beta)= \frac{e^{-\beta H}}{\mathcal{Z}(\beta)} \qquad \qquad \qquad  \mathcal{Z}(\beta) = \mathcal{Z}_1(\utilde{\lambda}_1) \qquad \qquad \utilde{\lambda}_1= 1 - \beta
\end{equation}
If we are also able to evaluate the fluctuations of the energy around the average $\me_1$, we can include the variance in the set of constraints. Given that we also know the average value of the energy, this is equivalent to specify the second moment $\utilde{\mathcal{C}}_2:=\Tr \rho H^2 - \me_2 = 0$. This gives the \emph{second-level ensemble} $\gamma_2$:
\begin{equation}
\gamma_2 =  e^{(\utilde{\lambda}_0 - 1)\mathbb{I}} e^{(\utilde{\lambda}_1 -1) H} e^{(\utilde{\lambda}_2 -1) H^2}  \qquad \qquad \qquad \qquad \utilde{\lambda}_0 = 1- \log \mathcal{Z}_2(\utilde{\lambda}_1,\Tlambda_2)
\end{equation}
where $\mathcal{Z}_2(\Tlambda_1,\Tlambda_2) = \Tr e^{(\Tlambda_1-1)H} e^{(\Tlambda_2-1)H^2}$ and the constraint equations $\utilde{\mathcal{C}}_1 = \utilde{\mathcal{C}}_2=0$ provide the following additional relations:
\begin{equation}
\frac{\partial \log \mathcal{Z}_2(\Tlambda_1,\Tlambda_2)}{\partial \Tlambda_1} = \me_1 \qquad \qquad \qquad \qquad \frac{\partial \log \mathcal{Z}_2(\Tlambda_1,\Tlambda_2)}{\partial \Tlambda_2} = \me_2
\end{equation}
Moreover, we also have the following consistencies equations:
\begin{equation}
\frac{\partial^2 \log \mathcal{Z}_2(\Tlambda_1,\Tlambda_2)}{(\partial \Tlambda_1)^2} = \frac{\partial \log \mathcal{Z}_2(\Tlambda_1,\Tlambda_2)}{\partial \Tlambda_2} = \me_2
\end{equation}
With respect to the thermal case, given by the first-level ensemble, the use of a Gaussian ensemble constitutes a novelty which will be explored elsewhere. We now write down the generic solution for the $n-$th level ensemble. In this case, the constrained optimization problem takes into account the normalization of the state and the first $n$ statistical moments $(\me_1,\me_2,\ldots,\me_{n})$ of the energy probability distribution $P_E$:
\begin{equation}
\gamma_n = \mathrm{Exp} \left[ \sum_{k=0}^{n} (\Tlambda_{k}-1)H^{k}\right] = \frac{\mathrm{Exp} \left[ \sum_{k=1}^{n} (\Tlambda_{k}-1)H^{k}\right]}{\mathcal{Z}_n(\Tlambda_1,\ldots,\Tlambda_n)}  \label{eq:ensemble}
\end{equation}
with
\begin{equation}
\mathcal{Z}_n(\Tlambda_1,\ldots,\Tlambda_n) :=\Tr \mathrm{Exp} \left[ \sum_{k=1}^{n} (\Tlambda_{k}-1)H^{k}\right]
\end{equation}
The presence of the constraints implies the existence of the following relations
\begin{equation}
 \frac{\partial \log \mathcal{Z}_n(\Tlambda_1,\ldots,\Tlambda_n) }{\partial \Tlambda_k} =\me_{k} \qquad \qquad  k = 1, \ldots, n
\end{equation}
which can be used to find the value of the Lagrange multipliers $(\Tlambda_1,\ldots,\Tlambda_n)$ as functions of the data $(\me_1,\me_2,\ldots,\me_{n})$.
Moreover, given the exponential form of the solution, we have the following consistencies equations:
\begin{equation}
 \frac{\partial^{p_1+\ldots+p_N} \log \mathcal{Z}_n(\Tlambda_1,\ldots,\Tlambda_n)  }{(\partial \Tlambda_{k_1})^{p_1} \ldots (\partial \Tlambda_{k_N})^{p_N}} = \me_{f(\{k_j\},\{p_j\})} \qquad \mathrm{where} \qquad f(\{k_j\},\{p_j\}):= \sum_{j=1}^N (k_j-1)^{p_j}  \leq n
\end{equation}
Whenever $f$ exceed $n$, the relation does not necessarily reproduce the higher moments of $P_E$. However, if it does, this is an indication that the $n-$th level ensemble provides a good approximation to the full diagonal ensemble $\rho_{\mathrm{DE}}$.

\section{Meaning of the approximation}

Intuitively, the hierarchy of ensembles $\left\{\gamma_n\right\}$ provides progressively better approximations to $\rho_{\mathrm{DE}}$. Here we make the statement more rigorous, highlighting three important features of these ensembles.
\begin{itemize}
\item First, since we are maximizing the entropy, adding constraints can not increase the optimal value:
\begin{equation}
S_{\mathrm{vN}}(\gamma_{n+1}) \leq S_{\mathrm{vN}}(\gamma_{n}) \,\,\,. \label{eq:entropy}
\end{equation}
Therefore, the different levels of the hierarchy have a specific order, which is set by the value of their von Neumann entropy:
\begin{equation}
\log D = S_{\mathrm{vN}}(\gamma_1) \geq S_{\mathrm{vN}}(\gamma_{2}) \geq \ldots \geq S_{\mathrm{vN}}(\gamma_{D-1}) \geq S_{\mathrm{vN}}(\gamma_D)= S_{\mathrm{vN}}(\rho_{\mathrm{DE}})
\end{equation}
\item Second, given that we are including progressively higher moments of the energy probability distribution $P_E$, the moment generating function $M_n(t)$ of the $n-$th level ensemble provides increasingly better approximations to the moment generating functions $M_{DE}(t)$ of the diagonal ensemble. These are defined as 
\begin{equation}
M_{DE}(t):= \Mv{e^{iHt}}_{DE} = \Tr \rho_{DE}e^{iHt} = \sum_{n=1}^D |c_n|^2 e^{iE_nt} \qquad \qquad M_n(t) := \Mv{e^{iHt}}_{\gamma_n} = \Tr \gamma_n e^{iHt}
\end{equation}
The Taylor series of $M_{n}(t)$ is 
\begin{align}
&M_{n}(t) = \sum_{l=1}^\infty \left. \frac{\partial^l M_{DE}(t)}{\partial t^l} \right\vert_{t=0} \,\,\, t^l = 1 + m^E_1 t + m^E_2 t^2 + \ldots + m^E_n t^n  + \frac{\partial^k M_n(t)}{\partial t^k} + \ldots
\end{align}
and the first $n$ derivatives of $M_n(t)$ are the same as the ones of $M_{DE}(t)$. For this reason, 
\begin{align}
&M_n(t) - M_{DE}(t) = \sum_{l=n+1}^\infty \left. \left(\frac{\partial^l M_n(t) }{\partial t^l} \right\vert_{t=0} - m^E_l \right) t^l
\end{align}
From the physical perspective, this is relevant to provide predictions about the out-of-equilibrium behaviour of the quantum system. Indeed, the moment-generating function $M_{DE}(t)$ is the fidelity of the state $\ket{\psi_t}$ at the time $t$ with the initial state: $M_{DE}(t)  = |\braket{\psi_0}{\psi_t}| := F(t)$. For pure states, the trace-distance $T(\rho,\sigma):=\frac{1}{2}\Tr \left[ \sqrt{(\rho-\sigma)^2}\right]$ reduces to a simple function of the fidelity: $T(\rho(t),\rho(0)) = \sqrt{1-F(t)^2}$. Hence, $F(t)$ evaluates how much the state at time $t$ becomes distinguishable from the initial state. We conclude that the approximation scheme proposed before is clearly able to catch the behaviour of the fidelity at small times, where only the first few derivatives (up to $n$) are relevant. \\

\item Third, we now prove that the $\gamma_n$ provide progressively better approximation to the diagonal ensemble $\rho_{DE}= \gamma_{D}$. This is relevant to make predictions, about the equilibrium physics, which go beyond the thermal ansatz. We note that, thanks to the exponential form of the $\gamma_n$ we have
\begin{equation}
S_\mathrm{vN}(\gamma_n) - S_\mathrm{vN}(\gamma_D) = D_{KL}\left( \gamma_D |\!| \gamma_n \right)
\end{equation}
where $D_{KL}(\rho|\!|\sigma)$ is the relative entropy $D_{KL}(\rho|\!|\sigma) := \Tr \rho \log \rho - \Tr \rho \log \sigma$. Together with Eq.(\ref{eq:entropy}), this means that
\begin{equation}
D_{KL}\left( \gamma_D |\!| \gamma_{n+1} \right) \leq D_{KL}\left( \gamma_D |\!| \gamma_{n} \right) \,\,.
\end{equation}
Therefore, the sequence $d_n:=D_{KL}(\gamma_D|\!|\gamma_n)$, for $n=1,\ldots, D$ converges monotonically to zero as $n$ approaches $D$. The relative entropy is undoubtedly an important quantity, as it provides a measure for the distinguishability of two quantum states. Despite that, it is not a metric. Hence, it does not provide a good definition of distance in the convex set of density matrices. Because of that, we resort to the trace-distance. In order to prove convergence to the predictions of the diagonal ensemble, we define the sequence $t_n := T(\gamma_n,\gamma_D)$, for $n = 1, \ldots, D$. The Pinsker Inequality provides an upper bound to the trace-distance of two quantum states which depends on the relative entropy:
\begin{align}
&T(\rho,\sigma) \leq \sqrt{\frac{1}{2}D_{KL}(\rho |\!| \sigma)}
\end{align}
Therefore, thanks to the Pinsker inequality and to the fact that $t_k \in [0,1]$, the sequence $t_k$ converges to zero as $n$ goes to $D$:
\begin{equation}
0 \leq t_k \leq \sqrt{\frac{1}{2}d_k} \quad \mathrm{and} \quad \lim_{k \to D} d_k = 0 \qquad  \Longrightarrow \qquad \lim_{k \to D} t_k = 0
\end{equation}
Even though we could not prove that the sequence $t_k$ is monotonic, the fact that it is upper-bounded by the monotonically decreasing sequence of the relative entropies $d_k$ is sufficient to conclude that \emph{as n increases, the $\gamma_n$ provide increasingly better approximations to the diagonal ensemble $\gamma_D=\rho_{DE}$}.\\

\end{itemize}





\section{Examples}
Now we discuss two simple examples of the proposed approximation scheme. We focus on the following Hamiltonian model 
\begin{align}
&H = \sum_{i=1}^L g \sigma_i^x + h \sigma_i^z + J \sum_{i=1}^{L-1} \sigma_i^z \sigma_{i+1}^z - J(\sigma_1+\sigma_L) \label{eq:Hamiltonian}
\end{align}
where $\sigma_{i}^{x,y,z}$ are the Pauli operator describing local magnetization at the $i$-th site. In particular, we focus on the following choice of the parameters: $g=0.9$, $h=0.75$, $J=1$ and we look at two system sizes $L=4$ and $L=10$, to show how the technique works. The values of the parameters are chosen, following Ref.\cite{Kim2013}, so that the level-spacing statistics of the Hamiltonian spectrum is robustly non-integrable. The initial state that we consider is always the anti-ferromagnetic state aligned along the $z$ direction: $\ket{\psi_0}=\ket{\uparrow ,\downarrow , \ldots}$. On the one hand, we diagonalize the Hamiltonian and compute the diagonal ensemble $\rho_{DE}$ via its energy probability distribution $p_{DE}(E_j) = |\braket{\psi_0}{E_j}|^2$. On the other hand, we can compute the various moments of $p_{DE}(E_j)$ without diagonalizing the Hamiltonian: $\mu_k^E = \bra{\psi_0}H^k\ket{\psi_0}$. The knowledge of the first $n$ moments is then used to set up and solve the constrained optimization problem, which yields the ensemble $\gamma_n$. 
\subsection{First example: L=4}
\begin{figure}[t]
\centering
\includegraphics[scale=0.39]{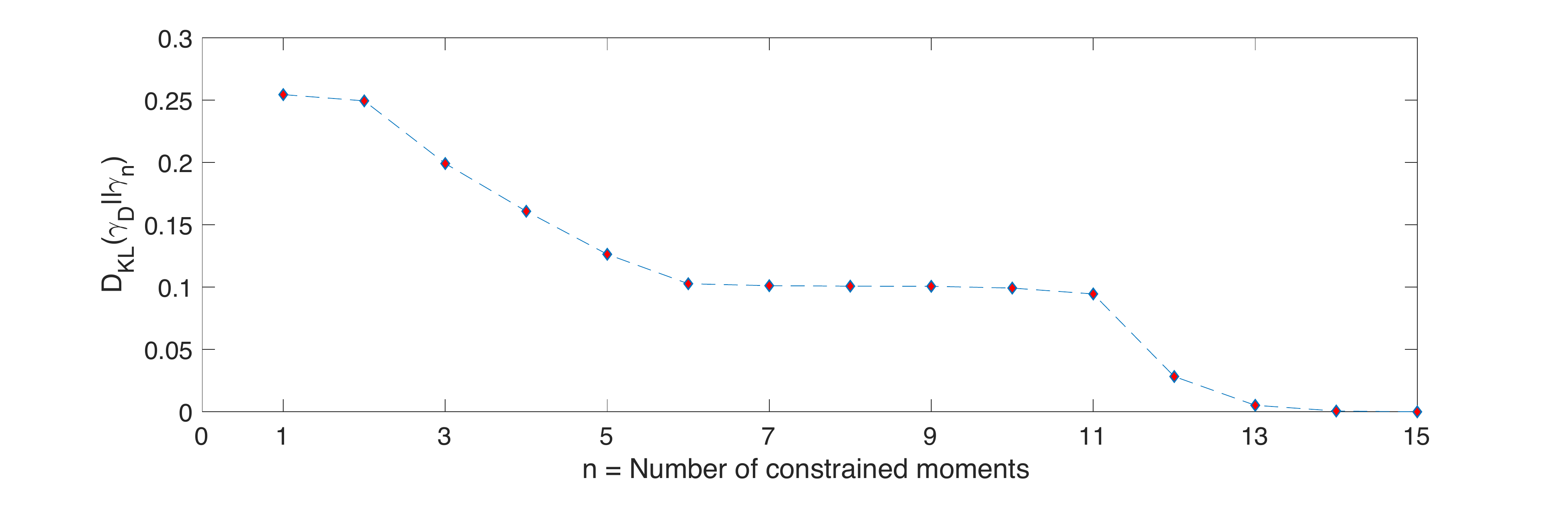}
\caption{Relative entropy $D_{KL}(\rho_{DE}|\!|\gamma_n)$ between $\gamma_n$ and $\rho_{DE}=\gamma_D$. The diagonal ensemble $\rho_{DE}$ is built from $\ket{\psi_0}=\ket{\uparrow ,\downarrow , \ldots}$ and the eigenstates of the Hamiltonian in Eq.(\ref{eq:Hamiltonian}). As $n$ increases we can see that $\gamma_n$ provides increasingly better approximations of $\rho_{DE}$.}\label{fig:DKL_Size4}
\end{figure}

\begin{figure}[h!]
\centering
\includegraphics[scale=0.45]{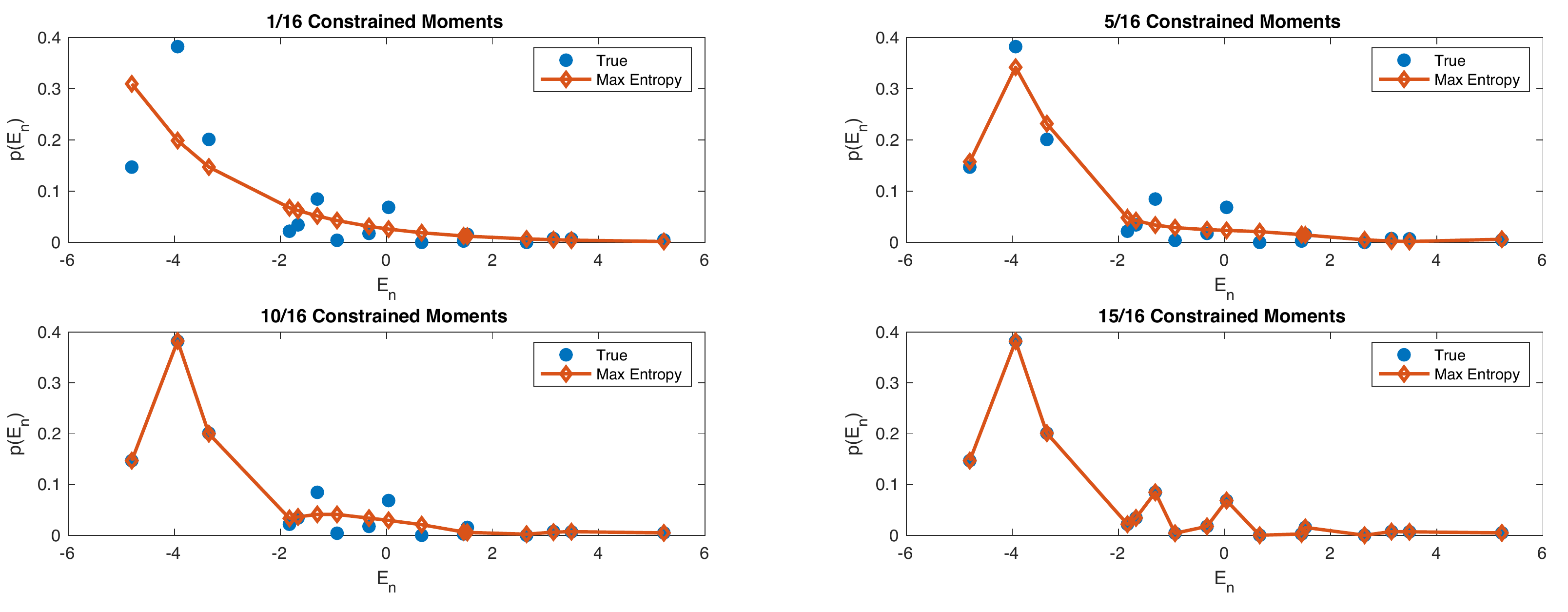}
\caption{Here we compare the shape of the true energy probability distribution (blue dots) with the maximum entropy distribution obtained with different numbers of constrained moments: 1(top left), 5 (top right), 10 (bottom left) and 15 (bottom right).}\label{fig:Shape_Size4}
\end{figure}

We start with the $L=4$ case, as it is simpler and it can be used to illustrate how the approximation scheme works. In Figure \ref{fig:DKL_Size4} we show the behaviour of the relative entropy $d_n=D_{KL}(\gamma_D|\!|\gamma_n)$ as $n$ increases. As proven in the previous section, this is monotonically decreasing and it becomes zero only when $n=D-1$. 
To get a sense of what is happening, in Figure \ref{fig:Shape_Size4} we plot the shape of the energy distribution computed from $\rho_{DE}$ and $\gamma_n$, at different values of $n$. It is clear that higher moments encodes the fine-grained details of the energy probability distribution. Thus, as we progressively constraint more moments, the $n$-th order ensemble captures more details of $\rho_{DE}$.
\subsection{Second example: L=10}
\begin{figure}[h!]
\centering
\includegraphics[scale=0.3]{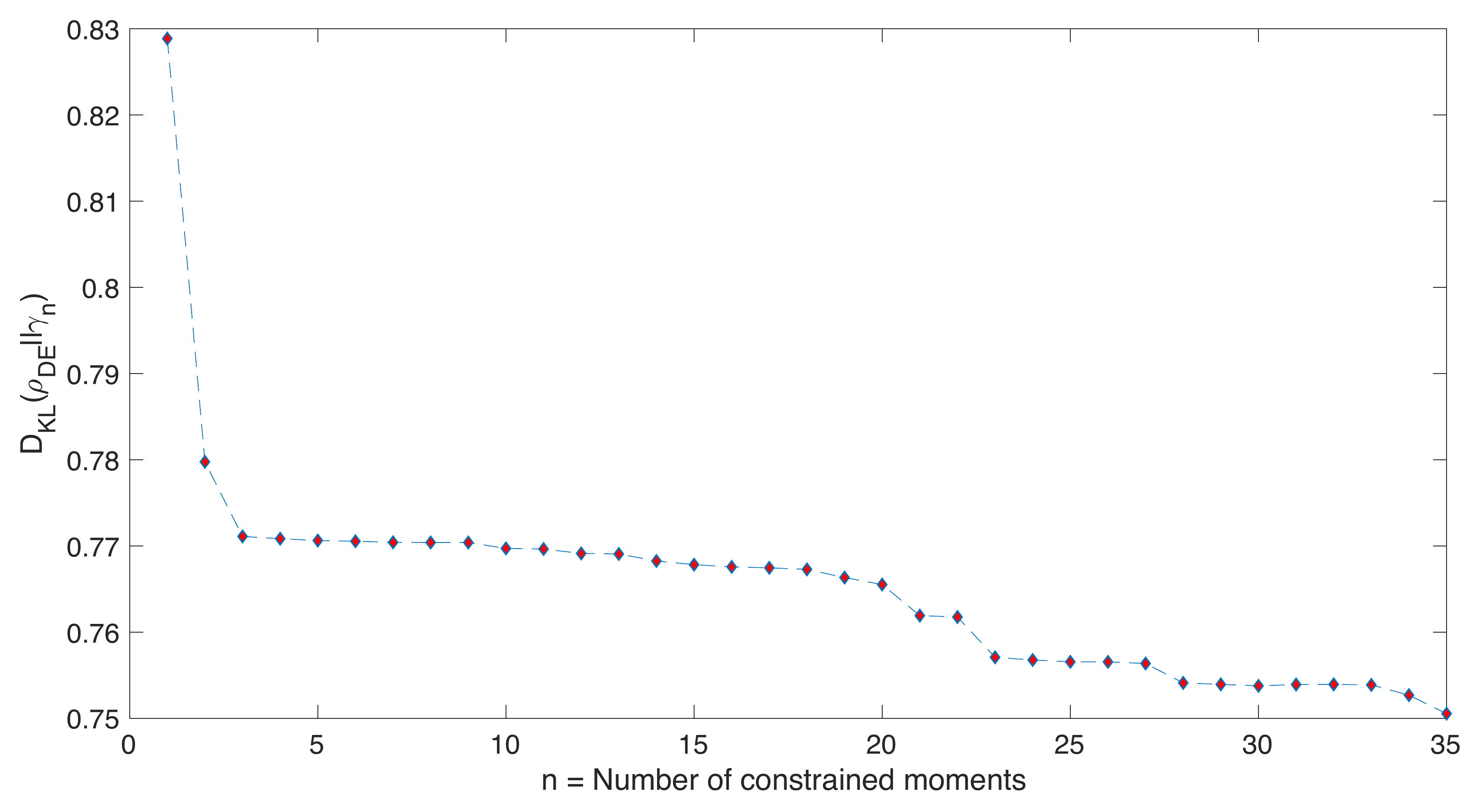}
\caption{Relative entropy $D_{KL}(\rho_{DE}|\!|\gamma_n)$ between $\gamma_n$ and $\rho_{DE}=\gamma_D$. The diagonal ensemble $\rho_{DE}$ is built from $\ket{\psi_0}=\ket{\uparrow_x ,\downarrow_x , \ldots}$ and the eigenstates of the Hamiltonian in Eq.(\ref{eq:Hamiltonian}) for system size $L=10$. As $n$ increases we can see that $\gamma_n$ provides increasingly better approximations of $\rho_{DE}$. However, we notice that only the first two moments provide a significant decrease in the relative entropy.}\label{fig:DKL_Size10}
\end{figure}

\begin{figure}[h!]
\centering
\includegraphics[scale=0.45]{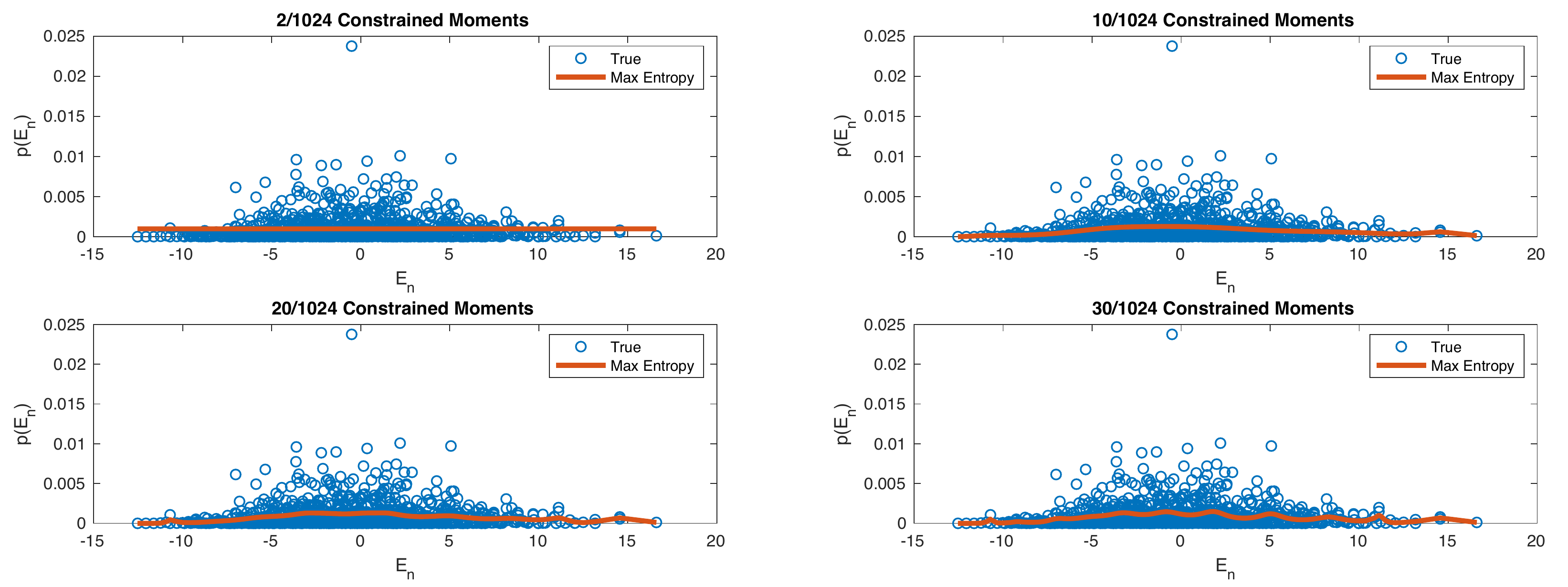}
\caption{Here we compare the shape of the true energy probability distribution (blue dots) with the maximum entropy distribution obtained with different numbers of constrained moments: 2 (top left), 10 (top right), 20 (bottom left) and 30 (bottom right).}\label{fig:Shape_Size10}
\end{figure}

Now we turn to the $L=10$ case, which is technically more involved, due to the fact that the dimensionality of the maximization problem is the same as the dimension of the Hilbert space, thus growing exponentially with the size of the system. Moreover, there is an upper limit about how many moments our computer is able to take into account. This is due to the fact that the value of the $n$-th moment is expected to grow exponentially with $n$. For example, considering that the spectrum of our Hamiltonian has boundaries which grow linearly with the size of the system $E_n \in (-\alpha L, \alpha L)$, with $\alpha \in \mathcal{O}(1)$, we might have to consider moments which are up to $n\in \mathcal{O}(2^L)$. For $L=10$ this means moments up to $n \sim 10^3$. The expected order of magnitude of these moments is $(\beta_n L)^{2^L}$ where $\beta_n$ is some constant $-1<\beta_n<1$. Therefore, irrespectively of what the concrete optimization algorithm is, there is a limit to the number of moments that we are able to take into account, given a certain size. Despite that, general informations about the shape of the distribution can always be obtained by taking into account as many moments as possible. In Figure \ref{fig:DKL_Size10} we plot the behaviour of the relative entropy between the diagonal ensemble $\rho_{DE}$, obtained considering as initial state the antiferromagnetic state along the $x$ direction $\ket{\psi_0^x}=\ket{\uparrow_x,\downarrow_x,\ldots}$, as the $n$-th ensemble $\gamma_n$.
Moreover, in order to understand how the entropy-maximization algorithms works, in Figure \ref{fig:Shape_Size10} we compare again the ``true'' energy probability distribution given by the diagonal ensemble $\rho_{DE}$ with the one given by the $n$-th ensemble, for increasing values of $n$. It becomes evident that, as we increase the size of the system, only the first two/three moments provide a significant amount of information about the whole probability distribution. This is witnessed by the fact that after $n=2,3$ we observe a neat plateaux in Figure \ref{fig:DKL_Size10}.

\section{Summary and conclusions}

In this paper we discuss a novel technique to address the equilibrium physics of isolated quantum system. The treatment builds upon Jaynes' derivation of statistical mechanics, through the maximum entropy principle. We have shown that, in order to obtain precise predictions about the equilibrium ensemble we need to know the full set of conserved quantities, which is given by the whole energy probability distribution $P_E$. Due to the fact that the dimension of the Hilbert space scales exponentially with the size of the system, this is highly impractical already for microscopic systems of modest size, as a chain of $N\sim 20$ qubits. Because of that, we explored the possibility to generate approximation schemes based on meaningful truncations of the full set of conserved quantities $P_E$.

We based our treatment on the fact that the most accurate solution, given by the diagonal ensemble $\rho_{DE}$, contains only constraints $\left\{ \mathcal{C}_n\right\}$ which are linear functionals. Because of that, we can use any (non-singular) linear combination of them $\utilde{\mathcal{C}}_k = \sum_n M_{kn} \mathcal{C}_n$ and the solution will be the same. Thinking in a geometric way, this can be understood as using different coordinates to describe the same entropy landscape. In the second part of the paper we studied a specific example: If we choose as matrix $M$ the Vandermonde matrix of the energy spectrum, $M_{nk}=(E_k)^{n-1}$, we see that specifying the full set of conserved quantities $P_E = \left\{ |c_n|^2\right\}$ is equivalent to specifying the $D$ statistical moments $\mu_j^E$ of $P_E$. As previously argued for the general case, this  suggests an approximation scheme, based on the experimental knowledge that we can gather about the energy of the quantum system under scrutiny. If the energy fluctuates so much that we can not even trust the average value, we use as constraint only the normalization of the state. This provides the zeroth-level ensemble, called $\gamma_0$. If the fluctuations around the average value are not too wild, the first moment provides meaningful information about the system and we can include it in the constraints. In this case we use the first-level ensemble, $\gamma_1$, which is equivalent to Gibbs Canonical Ensemble. Going further, if we are also able to evaluate properly the fluctuations, we can use the second-level ensemble $\gamma_2$, which has a Gaussian form. In general, if we have knowledge of the first $n$ statistical moments of the energy, plus the normalization of the state, we can use the $n$th-level ensemble $\gamma_n$ (see Eq.(\ref{eq:ensemble})) which is the exponential of an order-$n$ polynomial in the Hamiltonian.\\

The specific approximation scheme that we presented, based on the knowledge of the moments of the energy probability distribution, was inspired by the so-called Moments Problem\cite{Schmudgen2017,Mead1984,Tagliani2000} and the maximum entropy approach to tackle it. This technique is well-known within the Statistics and Information Theory community. Moreover, it has been previously employed in Nuclear Physics\cite{Weidenmuller2009,Behkami2018,Zelevinsky2016,Zelevinsky1996a,Senkov2013,Karampagia2018,Gomez2011,Koltun2010,Wong1986}, to model the density of energy levels (call it $g(E)$) in Hamiltonian models of nuclei. However, the analogy is purely technical as there is a key difference: Here we are trying to approximate a mixed state $\rho_{\mathrm{DE}}$, whose energy probability distribution is independent on the density of the energy levels. While $g(E)$ and $p(E_n)$ are often summoned together, for example in the thermodynamic limit, they are clearly independent on each other. On the one side, $g(E)$ is a feature of the Hamiltonian spectrum which does not depend on the initial state. On the other side, here we are interested in addressing $p(E_n) = |\braket{\psi_0}{E_n}|^2$, which is an initial-state-dependent quantity that specifies how much an energy eigenstates $\ket{E_n}$ is engaged during the time-evolution. Indeed, if one looks at the results presented in \cite{Senkov2014,Behkami2018,Senkov2013,Karampagia2018}, it appears that a Gaussian-like shape for the energy density is correct in several cases. Hence, to characterize the behaviour of $g(E)$ only the first two moments seem to be important. On the contrary, in Fig.\ref{fig:DKL_Size10} we see that this is not the case for $p(E_n)$. While we can see that the first two moments are ``more informative'' than others, we are still far away from having a relative entropy which is sufficiently small. This can be confirmed by looking at Fig.\ref{fig:Shape_Size10}, where we can see that if we take into account only the first two moments (as in the top-left panel) we loose a large amount of details about $p(E_n)$. Thus, while we believe that a correct model for $g(E)$ is an important step to understand the equilibrium properties of isolated quantum systems, especially in the large-system-size regime, this is conceptually separate from obtaining a decent approximation of the equilibrium ensemble $\rho_{DE}$, which is the problem we are tackling here.\\

The choice of using the knowledge of the moments of the energy probability distribution is not the only possible course of action. Indeed, in principle, the (non-singular) matrix $M$ is completely arbitrary. Therefore, a meaningful choice for $M$ should be driven by the physical properties of the system under scrutiny. For example, if we are dealing with a classically-integrable quantum system, which in the classical domain has an extensive number of local conserved quantities, there will be linear combinations of the energy eigenstates which provide conserved quantities which are local. In this case, the knowledge of such conserved quantities can be included in the set of constraints and it would give rise to the Generalised Gibbs Ensemble (GGE)\cite{Calabrese2016a,Vidmar2016,Vasseur2016,Caux2016,Ilievski2016}, which is currently being used to study the equilibrium properties of integrable quantum systems.

However, this is not the end of the story. It is clear from the treatment proposed in Section \ref{sec:Results} that, in the most general case, the GGE thus defined is only an approximation to the full diagonal ensemble $\rho_{DE}$. In this case, the approximation scheme is based on the degree of locality of the conserved quantities which have been included. Indeed, the actual number of conserved quantities is always exponentially large in the size of the system. In general, using an extensive number of them gives only an approximation which will work well as long as we are interested only in observables which are not ``too non-local''. \\

We conclude by mentioning two directions where we would like to expand the present work. First, the treatment proposed raises a fundamental question about the emergence of thermal equilibrium. Isolated quantum systems which exhibit thermal equilibrium clearly have the property that, in the thermodynamic limit, we only need the first one or two moments to make reliable predictions. Figure \ref{fig:DKL_Size10} goes along with this intuition as it shows that the lowest $2$ moments are the only ones which are ``really informative''. Their inclusion among the constraints lowers, significantly, the relative entropy. This is not true for moments higher than $2$, as we can see from the large plateaux in Fig.\ref{fig:DKL_Size10}. Since $T(\rho_{\mathrm{DE}},\gamma_n) \leq \sqrt{\frac{1}{2} D_{KL}(\rho_{\mathrm{DE}},\gamma_n)}$, the same argument can be applied to the trace distance $T(\rho_{\mathrm{DE}},\gamma_n)$, which evaluates our ability to tell apart two states. Because of that, to distinguish $\gamma_{2}$ from $\rho_{DE}$ is essentially as difficult as to distinguish $\gamma_{30}$ from $\rho_{DE}$. Hence, as far as distinguishability between $\gamma_n$ and $\rho_{\mathrm{DE}}$ is concerned, there is not much of a difference between $n=2$ and $n=30 \,\, \mathrm{or} \,\,35$. Thus, we focus on $\gamma_2$, which ha a Gaussian energy probability distribution with mean and variance fixed by the initial state. In the limit of large system size, thermodynamic consistency requires the energy fluctuations to be small. In concrete Hamiltonian systems it has been argued \cite{Biroli,Rigol2008} that, under very general conditions, the energy per particle has vanishing fluctuations in the thermodynamic limit: $\Delta E \sim \sqrt{N}$ so that $\frac{\Delta E}{N} \sim \frac{1}{\sqrt{N}} \ll 1$ when $N \gg 1$. This implies that, in the thermodynamic limit, the probability distribution $P_E$ can become so narrow that the first moment might be representative of the whole probability distribution. This argument is usually invoked in synergy with the Eigenstate Thermalization Hypothesis to argue for the emergence of microcanonical expectation values\cite{Anza2017,Anza2018a}. For the same reason, in the thermodynamic limit, the Gaussian-shaped energy probability distribution that we obtain from $\gamma_2$ can be so narrow that, also thanks to the action of the density of states $g(E)$, it effectively acts as a microcanonical probability distribution. Further studies to understand how such ``narrowing'' effect concretely occurs and how it is related to other approaches are certainly needed. 

Second, sufficiently small quantum systems can certainly escape the thermal equilibrium assumption. The framework developed here can be used to tackle their equilibrium (via the hierarchy of ensembles) and out-of-equilibrium (via the generating function of these ensembles) properties, given that the knowledge of a minimum number of statistical moments is available.

\vspace{6pt} 




\acknowledgments{F.A. acknowledges discussions with R. G. James and would like to thank J. Crutchfield and the Complexity Science Center for their hospitality during the final part of this work. Moreover, the author would also like to thank one of the Academic Editors, for pointing out the connection between the moments problem and the Nuclear Physics literature.}
\funding{The author would like to thank the ``Angelo Della Riccia'' foundation and St. Catherine's College of Oxford for their support to this research.}


\conflictsofinterest{The author declare no conflict of interest. The funding agencies had no role in the design of the study; in the collection, analyses, or interpretation of data; in the writing of the manuscript, and in the decision to publish the results.} 


\appendixtitles{no} 
\appendixsections{multiple} 
\appendix
\section{Constrained Entropy maximization}
To solve the constrained maximization problem we exploit the Lagrange Multipliers technique. In the first case, we have $D$ constraints, given by the set $P_E$ of conserved quantities. Hence, the set of constraints $\left\{\mathcal{C}_n = 0\right\}$ is simply: $\mathcal{C}_n := \Tr \rho \Pi_n - |c_n|^2$. Introducing a Lagrange multiplier $\lambda_n$ for each constraint $\mathcal{C}_n$, we define the auxiliary function $\Lambda (\rho,\left\{\lambda_n\right\}) = S_{\mathrm{vN}}(\rho) + \sum_{n=1}^D \lambda_n \mathcal{C}_n$, which can be freely optimized. 
\begin{equation}
\delta \Lambda = \frac{\delta S_{\mathrm{vN}}(\rho)}{\delta \rho} \delta \rho + \sum_{n=1}^D \mathcal{C}_n \,\, \delta \lambda_n + \lambda_n \frac{\delta \mathcal{C}_n}{\delta \rho} \delta \rho = 0
\end{equation}
Variation with respect to the Lagrange multipliers enforces the validity of the constraints. Variation with respect to the state $\rho$ provides the solution, as a function of the Lagrange multipliers. This can then be turned into a function of the data $P_E$, enclosed in the definition of the constraints. Given that 
\begin{align}
& \frac{\delta S_{\mathrm{vN}}(\rho)}{\delta \rho} = - \log \rho - \mathbb{I} &&\frac{\delta \mathcal{C}_n}{\delta \rho} = \Pi_n 
\end{align}
we obtain
\begin{align}
&\mathcal{C}_n = 0 && - \log \rho_{eq} - \mathbb{I} + \sum_{n=1}^D \lambda_n \Pi_n = 0 
\end{align}
This yields
\begin{align}
&\rho_{eq} = e^{- \mathbb{I} + \sum_{n=1}^D \lambda_n \Pi_n} &&\Tr \rho_{eq} \Pi_n = |c_n|^2 \label{eq:sol}
\end{align}
From the first equation, we can see that $\rho_{eq}$ must be diagonal in the energy eigenbasis. The second equation fixes these diagonal matrix elements to be $|c_n|^2$. Therefore, we obtain $\rho_{eq}= \rho_{\mathrm{DE}}$. Moreover, putting this form back into the first equation, we can find the value of the Lagrange multipliers. Indeed, using the properties of the projector $\Pi_n \Pi_m = \Pi_n \delta_{nm}$ and $= \mathbb{I} = \sum_{n=1}^D \Pi_n $, the first equation can be written as 
\begin{equation}
\rho_{eq}= e^{\sum_{n=1}^D (\lambda_n -1)\Pi_n} = e^{X} = \sum_{k=0}^{\infty} \frac{X^k}{k!}  = \sum_{n=1}^D \sum_{k=0}^{\infty} \frac{(\lambda_n-1)^k}{k!} \Pi_n = \sum_{n=1}^D e^{\lambda_n -1} \Pi_n
\end{equation}
Using the second part of Eq.(\ref{eq:sol}) we obtain $|c_n|^2 = e^{\lambda_n -1}$ and therefore $\lambda_n = 1 + \log |c_n|^2$. Here we can straightforwardly see how the Lagrange multipliers are connected to the data  $P_E$ enclosed in the constraint equations $\left\{\mathcal{C}_n = 0\right\}$.\\

Given the linear nature of the constraints $\left\{\mathcal{C}_n\right\}$, the solution of the optimization procedure is the same if we use linear combinations of them. In particular, using a vectorial notation for the Lagrange multipliers $\vec{\lambda} := \left\{ \lambda_n \right\}$ and the constraints $\vec{\mathcal{C}}:= \left\{ \mathcal{C}_n\right\}$ we have 
\begin{equation}
\vec{\utilde{\lambda}} \cdot \vec{\utilde{\mathcal{C}}} = \sum_{n} \utilde{\lambda}_n \utilde{\mathcal{C}}_n = \sum_{n,h,k} \lambda_k (M^{-1})_{kn} M_{nh} \mathcal{C}_h = \sum_{h,k} \lambda_k \delta_{kh} \mathcal{C}_h = \sum_{k} \lambda_k \mathcal{C}_k = \vec{\lambda} \cdot \vec{\mathcal{C}}
\end{equation}
where $\vec{\utilde{\mathcal{C}}}(\rho) :=M \vec{\mathcal{C}}$ or, $\utilde{\mathcal{C}}_n(\rho) :=\sum_{h}M_{nh}\mathcal{C}_h(\rho)$; $\vec{\utilde{\lambda}}:= \vec{\lambda}S^{-1}$, or $\utilde{\lambda}_n = \sum_{k}\lambda_k \left(S^{-1}\right)_{kn}$ and $M$ is a real non-singular matrix. In this way
\begin{equation}
\Lambda(\rho,\vec{\lambda}) = S_{\mathrm{vN}}(\rho) + \vec{\lambda} \cdot \vec{\mathcal{C}}(\rho) = S_{\mathrm{vN}}(\rho) + \vec{\utilde{\lambda}} \cdot \vec{\utilde{\mathcal{C}}}(\rho) 
\end{equation}
This proves that the solution of the optimization problem is the same, even though the specific form of the constraints and of the Lagrange multipliers is not. \\

\bibliography{library}



\end{document}